\documentclass[aps,pra,twocolumn,superscriptaddress,notitlepage]{revtex4-1}
\usepackage{amsmath,amssymb,amsthm,multirow,cases}
\usepackage{braket,dsfont,graphicx}
\usepackage[hidelinks,colorlinks=true,linkcolor=blue,urlcolor=blue,citecolor=blue]{hyperref}

\theoremstyle{remark}
\newtheorem*{subprotocol}{Subprotocol}
\DeclareMathOperator{\Tr}{Tr}
\usepackage{caption,subcaption}
\usepackage{wrapfig}
\usepackage{verbatim}

\begin{document}
\title{Multiparty quantum data hiding with enhanced security and remote deletion}
\author{Xingyao Wu}
\affiliation{Joint Center for Quantum Information and Computer Science, University of Maryland, College Park, MD 20742, US}
\author{Jianxin Chen}
\affiliation{Joint Center for Quantum Information and Computer Science, University of Maryland, College Park, MD 20742, US}

\begin{abstract}
One of the applications of quantum technology is to use quantum states and measurements to communicate which offers more reliable security promises. Quantum data hiding, which gives the source party the ability of sharing data among multiple receivers and revealing it at a later time depending on his/her will, is one of the promising information sharing schemes which may address practical security issues. In this work, we propose a novel quantum data hiding protocol. By concatenating different subprotocols which apply to rather symmetric hiding scenarios, we cover a variety of more general hiding scenarios. We provide the general requirements for constructing such protocols and give explicit examples of encoding states for five parties. We also proved the security of the protocol in sense that the achievable information by unauthorized operations asymptotically goes to zero. In addition, due to the capability of the sender to manipulate his/her subsystem, the sender is able to abort the protocol remotely at any time before he/she reveals the information.

\end{abstract}

\date{\today}


\maketitle


\section{Introduction}

Information security in the era of quantum information science has changed dramatically~\cite{valerio2009,acin2007,Gottesman2004}. With subfields like quantum cryptography~\cite{gisin2002,artur1991}, quantum teleportation~\cite{bennett1993,jianwei1997,ren2017ground}, quantum network~\cite{cirac1997quantum}, error correction codes~\cite{bennett1996mixed,cory1998experimental} etc., protocols for secure classical and quantum information sharing have been developed and, in some cases, implemented. While in the domain of classical steganography, applications have been made widely in covert message sharing~\cite{Bailey2006} and copyright protection~\cite{1369707}, one also seeks to bridge these information sharing schemes with quantum information since classical protocols are generally fragile in the sense that they depend on how well the senders keep the decoding strings and the existence of eavesdroppers would potentially undermine the security. Depending on whether the participants are allowed to have quantum communication, protocols are divided into two subcategories, namely quantum secret sharing (QSS)~\cite{hillery1999,xiaoli2004} where only local operations and classical communication (LOCC) are allowed and quantum data hiding (QDH)~\cite{terhal2001,divincenzo2002quantum} where quantum communication is allowed in certain ways. Experiments have also demonstrated the possibility of sharing secrets quantum systems~\cite{xiaoli2004}. 

In this work, we will be focusing on the part of quantum data hiding applied to multiple parties, i.e., more than two. Different from quantum secret sharing, the receivers in a quantum data hiding protocol are normally assumed to be individually malevolent which means they would do anything they could to decrypt the hidden information before the sender gives them the permission. Previous studies have been made with such two parties~\cite{terhal2001}. In that work, by utilizing the fact that the four Bell states cannot be distinguished by only LOCC, a classical information bit is securely encrypted in the parity of the number of the singlet states in a series of prepared Bell states. To later reveal the information, the two parties are given access to a shared quantum communication channel which enables them to distinguish the four Bell states easily.

Hiding information among multiple parties with quantum states has also been discussed in~\cite{eggeling2002,DiVincenzo2003}. However, these schemes will either require the allowed measurements to be in special forms or have very specific hiding structure considered. More general multipartite cases have been studied in~\cite{hayden2005}, but no example of the encoding states was given and the protocol success probabilistically. On the contrary, in this work, we will present a new multipartite quantum data hiding protocol which is equipped with new features mostly concerned with improvements of security compared to previous protocols~\cite{eggeling2002,DiVincenzo2003}, but also provides more straightforward measurements. 

The security of all the previous quantum data hiding protocols rely on the ability of the receivers to access certain quantum channels. The receivers are able to perform quantum nonlocal operations to reveal the hidden information given the authorization of the sender. We show that this may not be the safest option when hiding quantum data. Instead of relying on the authorization of the sender for the security, the protocol in this work enables the sender to hide the information bits securely even though the receivers maliciously get unauthorized access to quantum channel.

\section{The protocol}
\label{sec:protocol}

In a multipartite hiding scheme, comparing with the bipartite case, one important aspect we need to consider is how the receivers are collaborating with each other, specifically the grouping style of the receivers under which the information is hidden or revealed. For instance, given a set of recievers $\{A,B,C,D\}$, a partition of $\{\{A,B,C\},\{D\}\}$, where within the subgroups of $\{A,B,C\}$ or $\{D\}$ quantum communication is allowed and only classical communication between these two groups is allowed, will have different requirements on the protocol compared to a partition of $\{\{A,B\},\{C,D\}\}$. Thus, while considering the hiding protocol under the multipartite case, one also needs to take the hiding structure into account.

In a general setting, suppose a sender, denoted by $\mathcal{S}$, wants to share some bits of classical information to a group of receivers, denoted by $\mathcal{R}$. In the protocol, the sender wants to achieve the hiding and revealing of the bits only by modification (locally) on his/her part of the shared quantum state, while on the other hand, the receivers try their best to decrypt the hidden bits even though the sender has not given permission yet. 

Let us use the quantum state $\ket{\psi^b}$ to represent the hidden bits $b\in\{0,1,...,m-1\}$ which the sender want to share with the receivers. Unlike some typical protocols of data hiding proposed before~\cite{terhal2001,divincenzo2002quantum,eggeling2002,hayden2005}, the sender $\mathcal{S}$ here will still keep part of the system rather than sending states completely to the receivers. It will be shown later in section~\ref{security} and~\ref{erasetheinfo} that such treatment will actually improve the security in the hiding stage and also give the sender the flexibility to abort the protocol locally. The state on the receivers' side before the sender reveals or aborts will be
\begin{align}
    \rho^b_{\mathcal{R}_1\mathcal{R}_2...}=\Tr_\mathcal{S}[\ket{\psi^b}_{\mathcal{S}\mathcal{R}_1\mathcal{R}_2}...],
\end{align}
where $\mathcal{R}_i$ is the $i$th receiver.

In the hiding stage, the receivers are not able to guess the correct secret $b$. The sender can also specify how the receivers may collaborate with each other. For a partition $\mathcal{P}=\{\mathcal{P}_1,\mathcal{P}_2,...\}$, where $\mathcal{P}_j$ is the $j$th subgroup of the receivers among the set of the receivers $\mathcal{R}$, the sender will impose the restriction that only classical communication could be made across $\mathcal{P}_j$s and quantum communication could be made inside each group $\mathcal{P}_j$. The easiest way to realize these constraints is to make all the reduced states of the receivers to be the same, which is
\begin{align}
\label{rq1}
    \rho^b_{\mathcal{R}_1\mathcal{R}_2...}&=\rho^{b'}_{\mathcal{R}_1\mathcal{R}_2...},\\\nonumber
    (\textit{for}\quad b&\ne b')
\end{align}
This will guarantee that no matter what operation was made inside the group $\mathcal{P}_j$ and classical communication among $\mathcal{P}_j$s, the receivers cannot differentiate the encoding states for different $b$. We will show later in section~\ref{security} that the security in this hiding stage is unconditionally secure.

The next stage is to reveal the secret.
Since the sender still has part of the quantum state, he/she could manipulate the state in a way that will make the receivers able to recover the secret. This is done by making a local measurement on the sender's side. Suppose the positive-operator valued measure (POVM) elements corresponding to the sender's measurement are $\{M_i\}$, the state on the receivers' side after $\mathcal{S}$ made the measurement with outcome $t=0,...,m-1$ will be
\begin{align}
    \sigma^b_t=\frac{\Tr_{\mathcal{S}}[M_t\ket{\psi^b}\bra{\psi^b}M_t^\dagger]}{\Tr[M_t\ket{\psi^b}\bra{\psi^b}M_t^\dagger]}.
\end{align}

The sender will then inform the receivers classically with the outcome of the measurement, namely $t$. In this protocol, the sender would like the receivers in each partition $\mathcal{P}_j$ to collaborate nonlocally and receivers among different subgroups collaborate classically to recover the hidden information. To ensure this, we need to choose the encoding states carefully such that they satisfy these requirements. This simply means the states 
\begin{align}
\label{rq2}
    \Tr_{\cup_{k\ne j}\mathcal{P}_k}[\sigma^b_t],
\end{align}
with fixed $t$ are LOCC indistinguishable. To achieve this goal, we propose the following subprotocol.

\begin{subprotocol}
\textit{Suppose there are a set of multipartite states $\{\ket{\phi_l}\}$, where $|l|\geq m$, they satisfy a subprotocol if they cannot be perfectly distinguished by LOCC. The set of states $\{\ket{\phi_l}\}$ can also be divided into $m$ subsets and the $m$ states $\rho_{i}$ ($i=0,...,m-1$), where each of them is the uniform mixture of the states in each subset, also cannot be perfectly distinguished by LOCC.}
\end{subprotocol}

With the subprotocol defined above, for each subgroup $\mathcal{P}_j$, we can conveniently assign the state in~\eqref{rq2} to one of the uniform mixture $\rho_{b_j}$. Thus, in a subprotocol, the members in the subgroup $\mathcal{P}_j$ are attempting to recover the information $b_j$ with the allowed measurements. As stated already, only nonlocal measurement will enable each group $\mathcal{P}_j$ to perfectly distinguish the information $b_j$.

In order to force the receivers among different subgroup $\mathcal{P}_j$ to make classical communication to retrieve the hidden bits, the hidden bits $b$ could take the form
\begin{align}
    b=t\oplus b_1 \oplus b_2\oplus ...\oplus b_n \qquad(\textup{mod}\;m),
\end{align}
where $n$ is the total number of the subgroups and $t$ is the outcome of the POVM measurement made by the sender. In each round of the hiding protocol, the sender can decide a series $\{b_1,b_2,...,b_n\}$ which will encode the hidden bits $b$, and prepare the encoding state $\ket{\psi^b}$ with the right subprotocol states $\rho_{b_j}$. We will give examples of how to construct the subprotocol later in section~\ref{example}.

We have described above the protocol of hiding and revealing classical bits using quantum states. We will also discuss the security in the section~\ref{security}.

\section{A case study: hiding with $\{\{A,B\},\{C,D,E\}\}$}
\label{example}

\begin{figure}
    \centering
    \begin{subfigure}[b]{0.06\textwidth}
        \includegraphics[width=\textwidth]{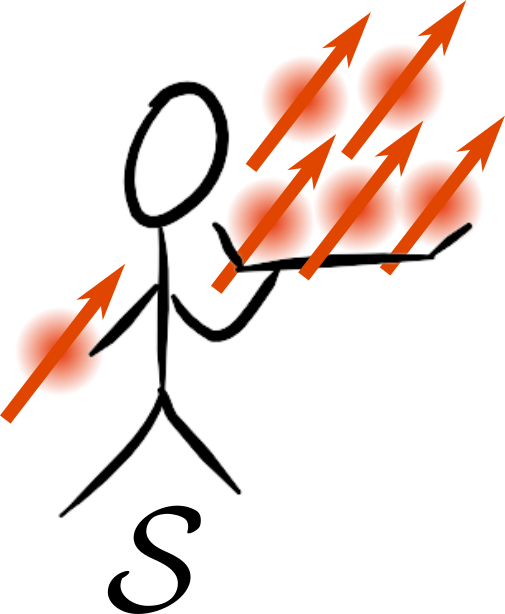}
        \caption{}
        \label{fig:prepare}
    \end{subfigure}\hspace{15mm}
    ~ 
    \begin{subfigure}[b]{0.2\textwidth}
        \includegraphics[width=\textwidth]{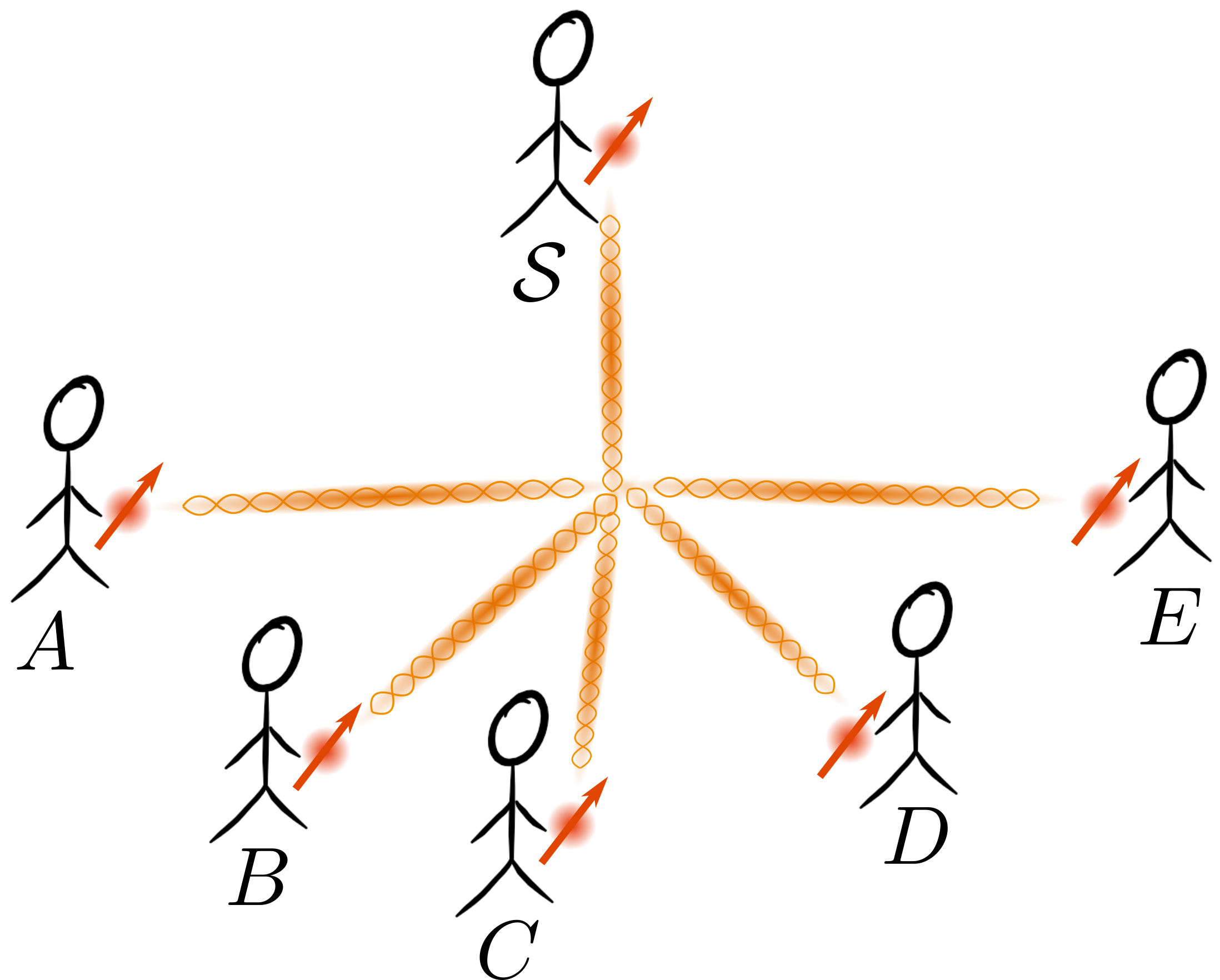}
        \caption{}
        \label{fig:send}
    \end{subfigure}
    \par\bigskip 
    \par\bigskip 
    ~ 
    \begin{subfigure}[b]{0.2\textwidth}
        \includegraphics[width=\textwidth]{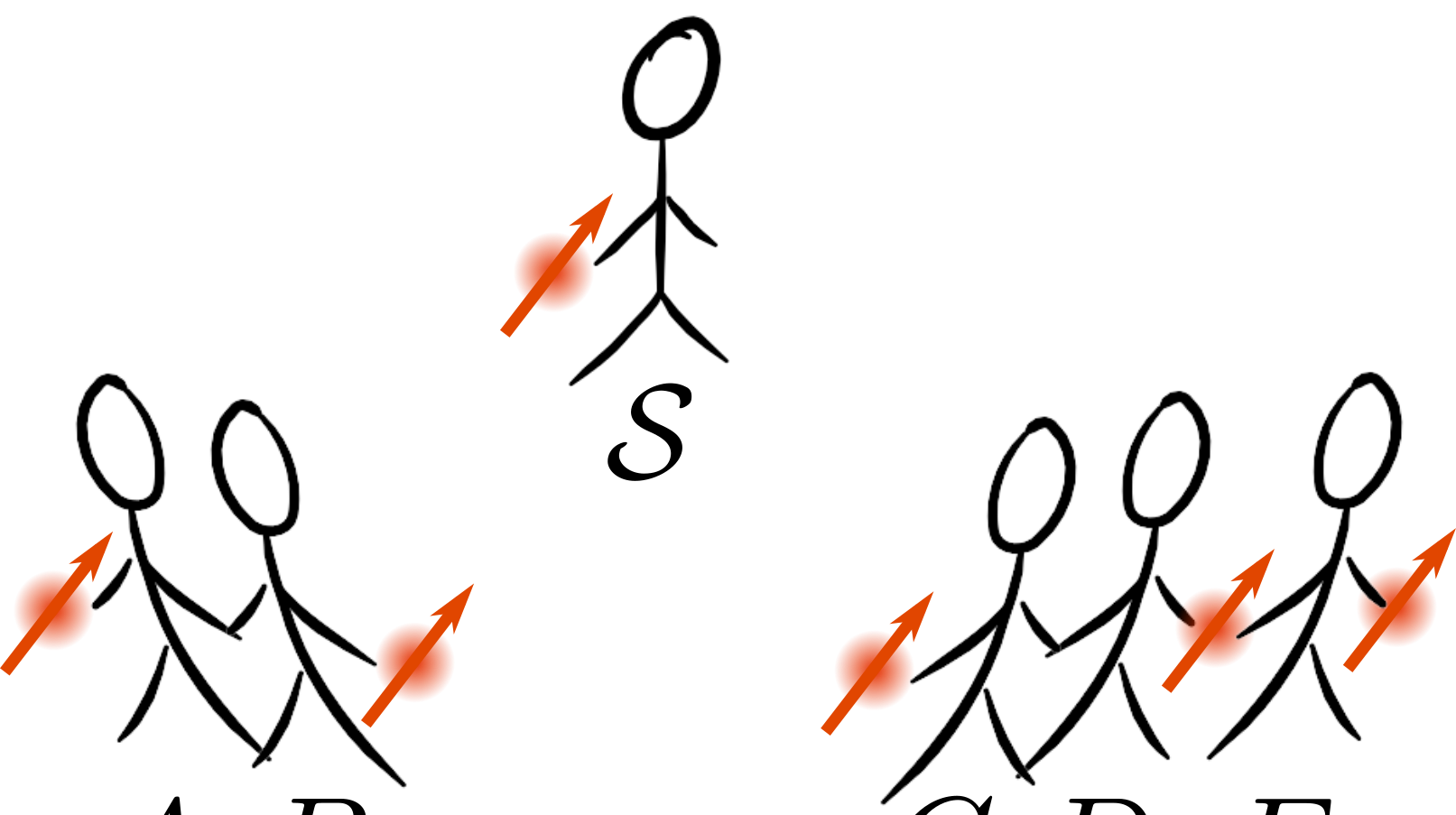}
        \caption{}
        \label{fig:hide}
    \end{subfigure}\hspace{7mm}
    \begin{subfigure}[b]{0.2\textwidth}
        \includegraphics[width=\textwidth]{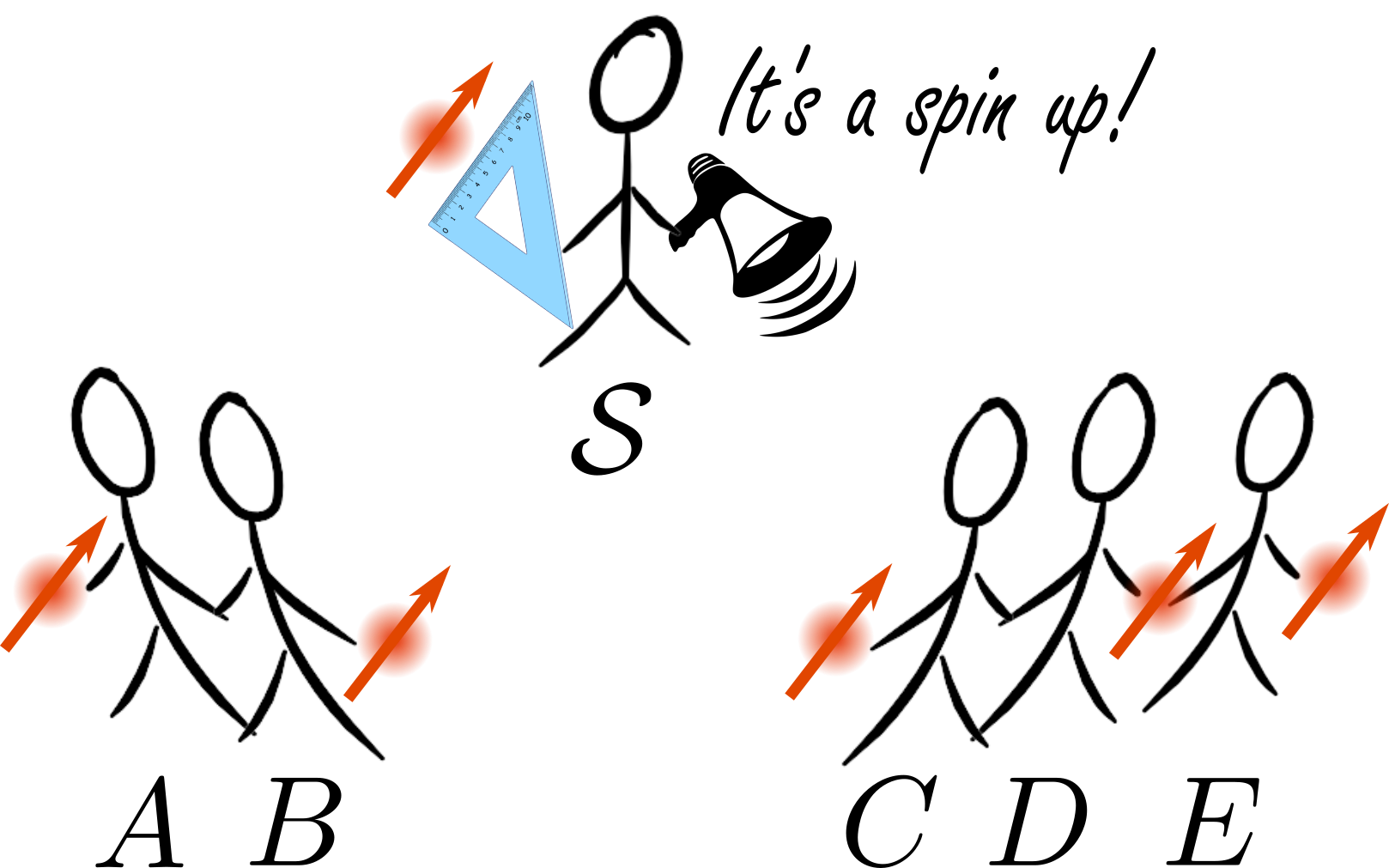}
        \caption{}
        \label{fig:reveal}
    \end{subfigure}
    \caption{An example case of the protocol when there are five receivers and the partition is $\{\{A,B\},\{C,D,E\}\}$. (\subref{fig:prepare}) The sender prepares the state in six subsystems. (\subref{fig:send}) The sender distributes the subsystems to the receivers. (\subref{fig:hide}) The receivers make two groups in the hiding stage. (\subref{fig:reveal}) The sender measures his/her qubit and announces the measurement result to the receivers enabling them to reveal the hidden information.}
    \label{protocol}
\end{figure}

In this section, we will show how could we construct the quantum data hiding protocol as we described above for a specific scenario, which has the receiver partition $\{\{A,B\},\{C,D,E\}\}$. Our protocol states that in the hiding stage, even though all the five members $\{A,B,C,D,E\}$ can make global quantum measurement, they still cannot recover the hidden bit (we will restrict to bit here and the generalization to multiple bits will be discussed at the end of the section). In the revealing stage, on the contrary, the sender will require the members in each subgroup to collaborate nonlocally to recover the hidden bit. Different stages of the protocol are depicted in Fig.~\ref{protocol}.

We will first describe how we choose the states of the subprotocol for the subgroup. Depending on the number of parties in each subgroup, we may have different subprotocols. For the convenience of experimental realization, we prefer that all the receivers will be given a qubit. We will show that, with this restriction, the subprotocol will be of different types for the first and second subgroups. The one we would like to adopt here for the first subgroup $\{A,B\}$ is the four Bell states protocol~\cite{terhal2001,divincenzo2002quantum}. Literally speaking, the four Bell states defined as $\ket{\Phi^{\pm}}=\frac{1}{\sqrt{2}}(\ket{00}\pm\ket{11})$ and $\ket{\Psi^{\pm}}=\frac{1}{\sqrt{2}}(\ket{01}\pm\ket{10})$, cannot be perfectly distinguished with an infinite number of LOCC 'rounds'\cite{divincenzo2002quantum}. 
Thus, the subprotocol states $\rho_{b_1}$ will be
\begin{align}
    \rho_{b_1}=\begin{cases}
    \frac{1}{2}(\ket{\Psi^-}\bra{\Psi^-}+\ket{\Phi^+}\bra{\Phi^+}), & \text{if $b_1=0$}.\\
    \frac{1}{2}(\ket{\Psi^+}\bra{\Psi^+}+\ket{\Phi^-}\bra{\Phi^-}), & \text{if $b_1=1$}.
  \end{cases}
\end{align}

On the other hand, for the second group $\{C,D,E\}$, we will choose the unextendible product bases (UPB)~\cite{Bennett1999,DiVincenzo2003} for the subprotocol states. It has been shown in~\cite{Rinaldis2004} that any set of states make up a UPB are LOCC indistinguishable. Actually, for a subprotocol of more than two parties, we can always define a UPB which involves only qubits and is LOCC indistinguishable~\cite{Augusiak2012,johnston2013} and thus could be used for the construction of the subprotocol. A neat and simple example of UPB states with three qubits are
\begin{align}
    \{\ket{000},\ket{1\Bar{e}e},\ket{e1\Bar{e}},\ket{\Bar{e}e1}\},
\end{align}
where $\{\ket{e},\ket{\Bar{e}}\}\in \mathbb{C}^{2}$ and the two states are orthogonal to each other but different than $\{\ket{0},\ket{1}\}$~\cite{Augusiak2012}. The subprotocol states $\rho_{b_2}$ can now be defined as
\begin{align}
    \rho_{b_2}=\begin{cases}
    \frac{1}{2}(\ket{000}\bra{000}+\ket{1\Bar{e}e}\bra{1\Bar{e}e}), & \text{if $b_2=0$}.\\
    \frac{1}{2}(\ket{e1\Bar{e}}\bra{e1\Bar{e}}+\ket{\Bar{e}e1}\bra{\Bar{e}e1}), & \text{if $b_2=1$}.
  \end{cases}
\end{align}

With the two subprotocol states ready, we can now write down the form of the encoding states $\ket{\psi^b}$. The state $\ket{\psi^0}$ will be chosen randomly from one of the following states
\begin{align}
    \ket{\psi^0}=\begin{cases}
    \ket{0}_{\mathcal{S}}\ket{0^*}_{AB}\ket{0^\#}_{CDE}+\ket{1}_{\mathcal{S}}\ket{0^*}_{AB}\ket{1^\#}_{CDE},\\
    \ket{0}_{\mathcal{S}}\ket{1^*}_{AB}\ket{1^\#}_{CDE}+\ket{1}_{\mathcal{S}}\ket{1^*}_{AB}\ket{0^\#}_{CDE},
    \end{cases}
\end{align}
where the register of the receivers are labelled by $b_1^*$ and $b_2^\#$, and $\ket{0^*}\in \{\ket{\Psi^-},\ket{\Phi^+}\}$, $\ket{1^*}\in \{\ket{\Psi^+},\ket{\Phi^-}\}$, $\ket{0^\#}\in \{\ket{000},\ket{1\Bar{e}e}\}$ and $\ket{1^\#}\in \{\ket{e1\Bar{e}},\ket{\Bar{e}e1}\}$, and moreover, they are all chosen randomly from each set.

Similarly, the state $\ket{\psi^1}$ will be chosen randomly from one of the following states
\begin{align}
    \ket{\psi^1}=\begin{cases}
    \ket{0}_{\mathcal{S}}\ket{0^*}_{AB}\ket{1^\#}_{CDE}+\ket{1}_{\mathcal{S}}\ket{0^*}_{AB}\ket{0^\#}_{CDE},\\
    \ket{0}_{\mathcal{S}}\ket{1^*}_{AB}\ket{0^\#}_{CDE}+\ket{1}_{\mathcal{S}}\ket{1^*}_{AB}\ket{1^\#}_{CDE},
    \end{cases}
\end{align}
where $\ket{b_1^*}$ and $\ket{b_2^\#}$ are chosen similarly as that for $\ket{\psi^0}$.

With the states defined above, it can be checked that they satisfy all the requirements we set for a quantum data hiding protocol described in section~\ref{sec:protocol}, and the corresponding measurement $\{M_i\}$ of the sender will be the projective measurement in the computational basis.

Regarding hiding multiple bits, it has been shown in~\cite{terhal2001,divincenzo2002quantum} that the sender can simply encode the bits into multiple parallel blocks of the above protocol. On the other hand, as the formalism of our protocol does not rely on the unique properties of the maximally entangled states as in~\cite{terhal2001,divincenzo2002quantum}, we do not have to restrict ourselves to qubits case and can go further into the qudit case for each of the parties. Since the UPB states are also available in higher dimensions~\cite{Wang2015,zhang2016}, to find proper states for the multiple bits is not a difficult task.

\section{Upper bound of the attainable information}
\label{security}

As suggested by the name of quantum data hiding, the main purpose of the protocol is that the sender can hide data among the receivers and then reveal the data only if the receivers later perform authorized operations. The security issue arises in two main aspects, namely, how well the protocol can hide the information bits and how much can the sender assure that the receivers are performing authorized operations.

We will first discuss the first aspect which is how well the protocol can hide the bits in the hiding stage. Since we assume the receivers to be malevolent, without loss of generality, we can assume the measurement made by the receivers to be a POVM $\{N_k\}$. The probability of the receivers to guess the bits as $i$ given the encoding states as $\ket{\psi^b}$ will be
\begin{align}
    p(b|k)=\Tr\Big[N_k\Tr_\mathcal{S}[\ket{\psi^{b}}_{\mathcal{S}\mathcal{R}_1\mathcal{R}_2}...]N^\dagger_k\Big].
\end{align}
With the promise of~\eqref{rq1} by the protocol, we know that $\Tr_\mathcal{S}[\ket{\psi^{b}}_{\mathcal{S}\mathcal{R}_1\mathcal{R}_2}...]$ is the same for all the values of $b$. Thus the above probability will be
\begin{align}
    p(b|k)=\frac{1}{m}.
\end{align}
Hence, in the hiding stage, the attainable information by performing the POVM $\{N_k\}$ will be the mutual information
\begin{align}
    I(B:K)&=\log_2 m-\sum_k p(k)H(B|K=k)\\\nonumber
    &=\log_2 m-m\times \frac{1}{m}\times\sum_b p(b|k)\log_2 p(b|k)\\\nonumber
    &=0,
\end{align}
where $B$ and $K$ corresponds to the events represented by the random variable $b$ and $k$. We have shown that the information attainable by the receivers is zero in the hiding stage. It should be stressed here that this is quite special compared to previous QDH protocols~\cite{terhal2001,eggeling2002,DiVincenzo2003,hayden2005}, since the attainable information there are all at a finite amount in the hiding stage. 

Now, we turn to the discussion about how much can the sender assure that the receivers are performing authorized operations. The idea is the same, which is to show how much information the receivers can attain by unauthorized operations, which will be LOCC operations in this case. Since the two subgroups are only allowed to have classical channel, we can assume the POVM elements for each of the group to be $\{P_i\}$, $\{Q_i\}$ and etc. To bound the attainable information Specifically, we want to bound the mutual information
\begin{align}
    &I(B;P,Q,...)=H(B)-H(B|P,Q,...)\nonumber\\
    \leq&H(B_1)-H(B_1|P)+H(B_2)-H(B_2|Q)+...\nonumber\\
    =&I(B_1;P)+I(B_2;Q)+...
\end{align}
The above equation simply tell us that the attainable information about $b$ is bounded by the sum of the attainable information of the subprotocols. Thus, one only needs to bound the attainable information of each subprtocol. For a good protocol, one always wants to bound this attainable information to zero. Here, we will use the example in section~\ref{example} to show how this is achieved.

In the case of hiding a bit, it involves a two outcome measurement for the subprotocol. Let us take the first subgroup $\{A,B\}$ as an example. It is shown in~\cite{divincenzo2002quantum} that if $|p(b_1=0|0)+p(b_1=0|1)-1|\leq \delta$, the mutual information between $B_1$ and $P$ will be bounded by 
\begin{align}
    I(B_1;P)\leq \delta H(B_1).
\end{align}
Since the two states in the subprotocol is not perfectly distinguishable by LOCC,
\begin{align}
    |p(b_1=0|0)+p(b_1=0|1)-1|< 1.
\end{align}
However, even if the value on the left hand side is smaller than one, it is definitely not going to bound the mutual information to be close to zero. To solve this problem, we consider the case where the subprotocol is composed with $n$ parallel original subprotocols and the answering bit $b_1$ is decided by the sum of the answering bit of each parallel session (modulo $2$). In this case, suppose $p(0|0)+p(0|1)=1+p$ for the original subprotocol,
\begin{align}
    p^{(n)}(0|0)+p^{(n)}(0|1)&\leq 2\sum_{i\;\text{even}}(1+p)^{n-i}p^i{{n}\choose{i}}\nonumber\\
    &=1+p^n.
\end{align}
Since $p\leq 1$ for all the subprotocols we chose, it will suffice to bound the attainable information to a desired small amount with large $n$. Thus, with enough resources, the sender can always make sure that the receivers may not recover the hidden bits with LOCC, which is not the authorized operation.

\section{Burn the message}
\label{erasetheinfo}

When hiding data in the real world, even though the sender could make his/her best to keep his/her subsystem in the safest place, it is still possible that the eavesdropper could make his/her way somehow to access it and use it for some malicious purposes. A simple example is that he/she could use it to send fake information to the receivers which is not what the sender wished them to see. The good thing of our protocol is that we allow the sender to abort the protocol at any time before the information was revealed. Let us say there is another POVM $\{K_i\}$ on the sender's side. Mathematically speaking, we require that the states in the protocol after the measurement will be indistinguishable for all $b$,
\begin{align}
    \frac{\Tr_{\mathcal{S}}[K_i\ket{\psi^b}\bra{\psi^b}K_i^\dagger]}{\Tr[K_i\ket{\psi^b}\bra{\psi^b}K_i^\dagger]}=e^{i\theta_{bb'}}\frac{\Tr_{\mathcal{S}}[K_i\ket{\psi^{b'}}\bra{\psi^{b'}}K_i^\dagger]}{\Tr[K_i\ket{\psi^{b'}}\bra{\psi^{b'}}K_i^\dagger]},
\end{align}
for $b\ne b'$ up to a global phase $e^{i\theta_{bb'}}$.

This means, immediately after $\mathcal{S}$ made the measurement, the quantum state shared by all the parties is erased to a state which cannot be used for information sharing anymore. With the example in section~\ref{example}, the corresponding measurement of the sender will just be the measurement $\{\ket{+}\bra{+},\ket{-}\bra{-}\}$. 

\section{Conclusion}
In this work, we presented a novel quantum data hiding protocol. The protocol extends the application scenario to arbitrary multipartite settings. We give the general requirements that the encoding states need to satisfy and illustrate how one could construct such a protocol with a specific example, which involves five parties. Since the states in the example are all pure qubit states, it is reasonable to say that the protocol is realizable with current experimental capabilities. In addition, due to the special construction of the encoding states, we proved that the protocol is promised to be unconditionally safe in the hiding stage, which is not the case for all previous protocols, and the sender can reduce the possibility of the receivers being cheating asymptotically to zero. Moreover, in section~\ref{erasetheinfo}, we show that it is possible for the sender to abort the data hiding protocol anytime before sending out the revealing permission.

As for the outlook for future study, even though the case for higher dimensional system is promised to be existing, we still need to find an explicit construction. Moreover, while we are parallelizing the subprotocol in section~\ref{example} to make the attainable information asymptotic zero, we assumed that the decoding strategy for the receivers cannot be better than measuring each block separately and then combine the results. Even though this is also assumed in previous works~\cite{divincenzo2002quantum}, we still look forward to a proof without such assumption.

\section*{Acknowledgement}

We acknowledge Eric Chitambar, Honghao Fu, Minh Tran and Alexey Gorshkov for helpful discussions. This work was supported by the ARL funded CDQI at the University of Maryland.

\appendix

\bibliography{IH}

\end{document}